\documentclass[12pt,preprint]{aastex} 

\slugcomment{Last Modified \today}

\shorttitle{Stellar Mass Estimation Based on IRAC Photometry}
\shortauthors{Zhu et al.}

\begin{document}

\title{Stellar Mass Estimation Based on IRAC Photometry for $Spitzer$ SWIRE-field Galaxies}

\author{Yi-Nan Zhu\altaffilmark{1,2,3}, Hong Wu\altaffilmark{1,2}, Hai-Ning Li\altaffilmark{1,2,3}, Chen Cao\altaffilmark{4}}
\altaffiltext{1}{National Astronomical Observatories, Chinese Academy of Sciences, Beijing 100012, China; zyn@bao.ac.cn; hwu@bao.ac.cn; lhn@bao.ac.cn}
\altaffiltext{2}{Key Laboratory of Optical Astronomy, National Astronomical Observatories, Chinese Academy of Sciences, Beijing 100012, China}
\altaffiltext{3}{Graduate University, Chinese Academy of Sciences, Beijing 100039, China}
\altaffiltext{4}{Institute of Space Science and Physics, Shandong University at Weihai, Weihai, Shandong 264209, China; ccao00@gmail.com}

\begin{abstract}
We analyze the feasibility to estimate the stellar mass
of galaxies by mid-infrared luminosities based on a large sample
of galaxies cross-identified from $Spitzer$ SWIRE fields and SDSS
spectrographic survey. We derived the formulae to calculate the
stellar mass by using IRAC 3.6$\mu$m and 4.5$\mu$m luminosities.
The mass-to-luminosity ratios of IRAC 3.6$\mu$m and 4.5$\mu$m
luminosities are more sensitive to star formation history of
galaxies than other factors, such as the intrinsic extinction,
metallicity and star formation rate. To remove the affection from
star formation history, we used g-r color to recalibrate the
formulae and obtain a better result.  It must be more careful to
estimate the stellar mass of low metallicity galaxies using our
formulae. Due to the emission from dust heated by hottest young
stars, luminous infrared galaxies present higher IRAC 4.5 $\mu$m
luminosity compared to IRAC 3.6 $\mu$m luminosity. For most of
type-II AGNs, the nuclear activity can not enhance 3.6$\mu$m and
4.5$\mu$m luminosities compared with normal galaxies. The star
formation in our AGN-hosting galaxies is also very weak, almost
all of which are early-type galaxies.
\end{abstract}


\section{Introduction}

It is important to obtain the mass of galaxies for understanding the evolution of the universe. We know that the dark matter halo contains
most of mass, and can only be detected using the effect of gravitation. So generally, the total mass
of galaxies could only be computed on the basis of kinematics of bright stars, clusters, or even satellite galaxies \citep[e.g.,][]{rogstad72, roberts73, ostriker74}.
However, due to the restriction from lower resolution and lower sensibility of telescopes for detecting the motion of celestial bodies
in high redshift galaxies, these methods are just used for deriving general properties of nearby galaxies. Though the fraction of baryonic
components in galaxies is relatively small, it could be detected and studied easily. Except the gas dominant galaxies,
stars hold most of baryonic components in galaxies. Thus, to understand the correlation between the stellar mass
and total mass is essential for studying the faint and distant galaxies. Therefore, it is necessary to seek some methods to estimate
the stellar mass in galaxies.

If we know the mass-to-luminosity ratio of a galaxy, and assume an
initial mass function (IMF), its stellar mass could also be
estimated from the luminosity in corresponding wavelength
\citep[e.g.,][]{bell01, portinari04}. Traditionally,
mass-to-luminosity ratios could be derived by fitting photometric
colors or spectra with models. Most of baryonic components are
trapped in low mass stars, which dominate the output of galaxies
in longer wavelength range compared to the massive ones. At the
same time, the short life scale of massive stars and heavy
extinction in relatively short wavelength ranges restrict us to
use them to detect the bulk of galactic stellar masses. Hence,
less affected by other factors, such as metallicity, star
formation rate (SFR) and star formation history (SFH),
luminosities in relatively longer wavelength (for example,
near-infrared (NIR)  band), are ideal tracers of the stellar mass
\citep[e.g.,][]{cole01,kochanek01}. The $K$ band luminosity
distribution of distant galaxies was discussed in support of the
hierarchical picture by \citet{kauffmann98}. The main uncertainty
in the inferred stellar masses arises from the age of the stellar
population \citep{rix93}, similar to the results calculated by
optical luminosities \citep{bell05}. The discrepancy among stellar
masses derived from various methods has been found
\citep[e.g.,][]{drory04}. For example, the uncertainty of the
stellar mass estimated by $K$-band luminosity alone has been
discussed by \citet{brinchmann00}. \citet{kannappan07} compared
many estimation methods for local galaxies and also showed the
differences (the factors up to ~2) among them.

Except for above methods, the stellar mass could also be directly
calculated using mid-infrared (MIR) luminosities in 3-5$\mu$m,
since emissions from the photosphere of old stars dominate the
output of galaxies in this wavelength range. At the same time, the
extinction and reddening in this range are clearly weak. But using
these emissions to study the universe was prevented because of the
absorption of atmosphere until the onset of space infrared
observations. $Spitzer~Space~Telescope$ \citep{werner04} is a very
useful facility to help us to study the MIR to far-infrared (FIR)
emission properties of galaxies. The two detectors in the shorter
wavelength bands (3.6$\mu$m and 4.5$\mu$m) of $Spitzer$ Infrared
Array Camera \citep[IRAC;][]{fazio04} could be treated as stellar
mass tracers of nearby galaxies, although some disturbances may
exist, such as the continuum from hot dust or spectral features
from the polycyclic aromatic hydrocarbons \citep[PAHs;][]{leger84,
puget89, draine03, wu05}. Using about 150 local galaxies,
\citet{li07} showed that there truly existed tight correlations
between the K-corrected 3.6$\mu$m
luminosities and the stellar mass based on the  Bell et al.'s (2003) formula.

In this paper, we want to derive the stellar mass formulae using IRAC 3.6 and 4.5 $\mu$m luminosities based on the observation
of $Spitzer$ and Sloan Digital Sky Survey \citep[SDSS;][]{york00}. The $Spitzer$ Wide-area Infrared Extragalactic Survey \citep[SWIRE;][]{lonsdale03},
with a total field of $\sim$49 deg$^{2}$, is the largest extragalactic survey program among the six $Spitzer$ cycle-1 Legacy Programs,
and provides us a best opportunity to establish such correlations.

The structure of the paper is as follows. We describe the construction of our sample and the referenced stellar mass in $\S$2.
The major results of the MIR stellar mass estimations are presented in $\S$3. Discussion and summary are given
in $\S$4 and $\S$5. Throughout this paper, we adopt a $\Lambda$CDM cosmology with $\Omega_{\rm m}=0.3$, $\Omega_{\rm \Lambda}=0.7$
and $H_{\rm 0}=70\,{\rm km \, s^{-1}Mpc^{-1}}$.

\section{The Sample and Reference Stellar Mass}

The optical spectral sample galaxies are selected from the Sloan Digital Sky Survey \citep[SDSS;][]{york00} main galaxy sample
\citep{strauss02}, with $r$-band Petrosian magnitudes lower than 17.77 mag. A total $15$ deg$^{2}$ of SDSS sky matches the three
northern $Spitzer$ SWIRE fields. The derived data from SDSS DR7, including the stellar mass, was supplied by The Max-Planck-Institute
for Astrophysics ($MPA$) and the Johns Hopkins University ($JHU$) in their archives \footnote{http://www.mpa-garching.mpg.de/SDSS/DR7}.
Compared with former data releases, such as DR4, some galaxies were excluded in DR7 due to new restricts on the detected
sources before publication by $MPA$/$JHU$.
The stellar mass derived from fitting photometries with population synthesis models was treated as reference.
The foreground extinction in all the five SDSS bands were corrected by subtracting
the extinction values presented in the SDSS catalog. The IDL code by Blanton (version tag v4\_1\_4) was used to calculate the K-correction.
The method and the SED used in this code were described in details by \citet{blanton03} and \citet{blanton07}.

The IRAC four bands (3.6, 4.5, 5.8 and 8.0~$\mu$m) images were
mosaicked from the Basic Calibrated Data (BCD; acquired from the
$Spitzer$ Sciences Center) after the flat-field corrections, dark
subtraction, linearity and flux calibrations
\citep{fazio04,huang04}, with the final pixel scale of
0.6$\arcsec$ \citep{wu05,cao07,cao08}. The MIPS 24$\mu$m images
were mosaicked as the same way but with a pixel scale of
1.225$\arcsec$ \citep{wen07, cao07, wu07}. Based on the catalogs
of the Two Micron All Sky Survey
\citep[2MASS;][]{skrutskie97,cutri03}, we enhanced the accuracy of
the astrometric calibration of 0.1$\arcsec$ in all five bands. The
final IRAC and MIPS 24~$\mu$m flux has calibration uncertainties
less than 10\% \citep{rieke04}. Then we matched these MIR sources
with the above SDSS sample galaxies with a cross radius of
2$\arcsec$. For the purpose of this work, the sources with both
IRAC 3.6$\mu$m and 4.5$\mu$m detections were selected. The basic
sample contains $1454$ objects. For all these sources, a set of
Spectral Energy Distribution (SED) \citep{huang07} was used to
perform K-correction in IRAC 3.6$\mu$m and 4.5$\mu$m bands.

Only part of the galaxies (total is 659) have been mapped by $2MASS$.
The Extended Source Catalog ($XSC$) was download from the archives of Infrared Processing and Analysis Center [$IPAC$] \footnote{ http://irsa.ipac.caltech.edu/}
and matched with the above sample with a cross radius of 2$\arcsec$. The K-correction for $2MASS$ NIR band flux was also based on
the IDL code by \citet{blanton03} and \citet{blanton07}. The difference in $K$$_{\rm s}$ band between Vega magnitude which was used
by $2MASS$ catalogs and AB magnitude is 1.84 \citep{finlator00}. And a factor of 5.12 \citep{finlator00, binney98}was used as the $K$$_{\rm s}$ band solar absolute magnitude.

\subsection{Spectral Classification}
\begin{figure}
\figurenum{1} \epsscale{} \plotone{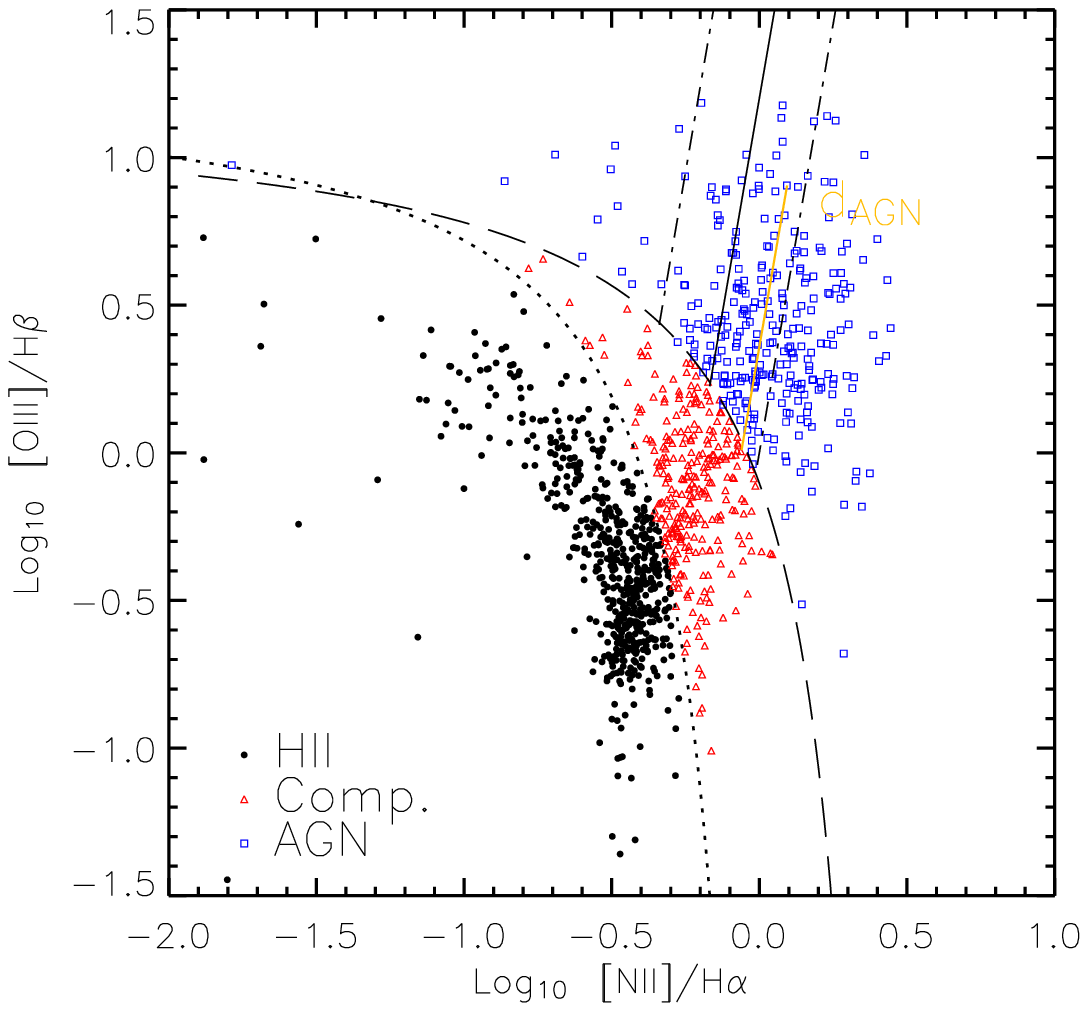}
\caption{The BPT
diagnostic diagram: [NII]/H$\alpha$ vs [OIII]/H$\beta$. The criteria from \citet{kauffmann03b} and \citet{kewley01} are illustrated
as dotted and dashed curves, respectively. The objects below the dotted curve are defined as star-forming galaxies. The triangles
between the above two curves are those classified as composite galaxies, while the boxes above the dashed curve denote AGNs.
We use the distance $d_{AGN}$, which was defined by \citet{wu07}, to characterize AGN activity. The two dotted-dashed lines mark the boundaries that were used by \citet{wu07}. An example was show in this figure
with orange solid line.}
\label{fig1}
\end{figure}

For all the 1454 galaxies, 68 with positive H$\alpha$ equivalent width were classified as absorption line galaxies,
while the rest 1386 were classified as emission line galaxies. Futhermore, emission line galaxies with detected H$\alpha$, H$\beta$,
[OIII], [NII] emission lines were selected, and optical spectral classifications were carried out on these  $1152$ galaxies,
adopting the traditional BPT diagnostic diagram: [NII]/H$\alpha$ versus [OIII]/H$\beta$ \citep{baldwin81, veilleux87}, as shown
in Figure~\ref{fig1}. The dashed curve is from \citet{kauffmann03b} and the dotted curve is from \citet{kewley01}. Objects located
below the dashed curve were classified as star-forming galaxies; those between these two lines were classified as composite (starburst
$+$ AGN) galaxies \citep{kewley06,wu98}; while those above the dotted line were classified as narrow-line AGNs. Therefore, the
final sample contains $1220$ galaxies, including 561 star-forming galaxies, 292 composites, 299 narrow-line AGNs, and 68 absorption
line galaxies, and $544$ of this final sample have the $2MASS$ $K$$_{\rm s}$ band flux.

\citet{wu07} presented an equation to compute $d_{AGN}$ to represent AGN activities, defined as the distance
of an AGN from its position ($x_p$,$y_p$) in dex in the traditionally line-diagnostic diagram [OIII]/H${\beta}$ versus [NII]/H${\alpha}$
to the \citet{kewley01}'s curve along the parallel of the best linear fitting of 27 AGNs. This is quite similar to the star-forming
distance defined by \citet{kewley06} in the diagnostic diagram [OIII]/H$ {\beta}$ versus [OI]/H${\alpha}$ or [SII]/H${\alpha}$.
Here, we also use this formula to quantify AGN activities (see the orange solid line in Figure~\ref{fig1} as an example). In Figure~\ref{fig1},
the two dotted-dashed lines above the dotted curve marked the boundaries that were used by \citet{wu07}. Only sources whose
EW(H$\beta$) were less than -5 were selected by \citet{wu07}. But in this work, we did not use this limit to reduce
those sources with smaller values of EW(H$\beta$), such as LINERs. Therefore, many AGNs appeared outside the right boundary in Figure~\ref{fig1}.

\subsection{Define Referenced Stellar Mass}

The thermally pulsing asymptotic giant branch (TP-AGB) could
dominate the output of galaxies at some evolutionary phases
\citep{maraston06, bruzual07}. But lots of evolutionary tracks
used in stellar population synthesis models did not consider the
effect from TP-AGB. So there must exist bias for the derived
stellar mass when comparing observations with those models. For
example, for young template models (with ages less than 2 Gyr),
both \citet{maraston06} and \citet{bruzual07} have concluded that
galactic stellar masses calculated with an improved treatment of
TP-AGB stars are roughly 50 to 60 per cent lower. Hence, due to
the lower effective surface temperature of TP-AGB, referenced
stellar mass derived from optical observations could avoid the
emission of TP-AGB.

The referenced stellar mass used in this work was provided by
$MPA$/$JHU$ in their archive
\footnote{http://www.mpa-garching.mpg.de/SDSS/}. They derived
stellar masses from fitting photometries \citep{salim07} and
spectral features \citep{gallazzi05,kauffmann03a} with a large
grid of models from BC03 \citep{bruzual03} spanning a large range
in SFH. The stellar mass derived using SDSS photometries could
avoid the aperture effect which should be considered when fitting
spectral features, since only a part of region of the observed
object was covered by SDSS spectral fiber with 3$\arcsec$
diameter. For about eighteen thousand galaxies randomly selected
from MPA SDSS DR4 Sample, the stellar masses derived from
photometries and spectral indices fits were compared and shown in
Figure~\ref{fig2}(a). We found a good correlation between them.

If the luminosity in some wavelength ranges could be obtained, we
can estimate the stellar mass by assuming the mass-to-luminosity
ratio in corresponding wavelength range. \citet{bell03} presented
the relationship between various optical colors and
mass-to-luminosity ratios. In Figure~\ref{fig2}(b), we compared
the referenced stellar mass with those computed based on the
equation by \citet{bell03}. Here, to correct the 'diet Salpeter'
IMF \citep{bell03} to the normal Salpeter \citep{salpeter55} IMF,
we added a factor of 0.15 \citep{bell03} to the stellar mass
calculated by using rest-frame $g$-$r$ color and $r$ band model
magnitude. And a factor of 4.67 \citep{bell03} was treated as the
$r$ band solar absolute magnitude (note there is a difference with
a factor of 0.09 against the value used by \citet{blanton03}). For
the stellar mass derived from the equations of \citet{bell03},
there is a systematical overestimation with factor of about
~0.2$-$0.6 dex, and such discrepancy is more obvious for a galaxy
with lower stellar mass. This trend is more clear in
Figure~\ref{fig3}. The disagreement is more obvious for lower
stellar mass and bluer galaxies than higher stellar mass and
redder ones, which is similar to the results supplied by of
\citet{kannappan07} who found that large discrepancy existed
between their reference stellar mass and the one derived from Bell
et al.'s (2003) correlations, with factors in the range of 2 to 5.

\subsection{Sample Distribution}

The distributions of the SDSS B-band absolute magnitude (M$_{\rm B}$), redshift, referenced stellar mass and $r$ band concentration
parameters of star-forming galaxies, composite galaxies, AGNs and absorption line galaxies are shown in Figure~\ref{fig4}. Here,
M$_{\rm B}$ were calculated from the SDSS $g$ and $r$-band model magnitudes according to \citet{smith02}. Most of the galaxies, especially
for composites, AGNs and absorption line galaxies, have M$_{\rm B}$ brighter than $-$18 mag, which usually is regarded
as the boundary to distinguish with dwarf galaxies \citep{thuan81}. Few galaxies have redshift larger than 0.15.
Herce, most of the galaxies in our sample are local ones. The star-forming galaxies have lower stellar mass than the others in Figure~\ref{fig4}(c).
The morphology of galaxies was represented by $r$ band concentration parameters ($R$50/$R$90), which defined as ratio of two radii containing 50\% and 90\% of the Petrosian $r$ band luminosity.
The values of $R$50/$R$90 are smaller for early-type galaxies than those of late-type galaxies. According the distributions of four
different spectral type galaxies in Figure~\ref{fig4}(d), we found that the morphologies of the composite galaxies, AGNs and absorption
line galaxies are similar, but those of star-forming galaxies are later type. Therefore, on the basis of above discussions,
we find that low luminous dwarf early-type galaxies are scarce in this sample.

\section{Result: Stellar Mass Estimated by MIR luminosities}

A lot of factors could effect the stellar mass estimations.
Therefore, good stellar mass tracers should be insensitive to
these factors. Figure~\ref{fig5} show the correlations between
mass-to-luminosity ratios of $2MASS$ $K$$_{\rm s}$ band and two
MIR bands luminosities and various factors. The $K$$_{\rm s}$
luminosity was proved to be an ideal stellar mass tracer.
Figure~\ref{fig5}(a1, a2, a3) present the effect of EW(H$\alpha$)
on the mass-to-luminosity ratios. EW(H$\alpha$) is H$\alpha$
equivalent width and could be regarded as the representative of
SFH \citep{kennicutt98}. The concentration parameters
($R$50/$R$90) in Figure~\ref{fig5}(b1, b2, b3) represent the
morphology of galaxies. The effect of color excess E(B-V) on the
mass-to-luminosity ratios is illustrated in Figure~\ref{fig5}(c1,
c2, c3). E(B-V) is for the intrinsic extinction and calculated
from Balmer decrement $F_{\rm H\alpha}/F_{\rm H\beta}$
\citep{calzetti01}. Thus only the emission line galaxies are
plotted here. In Figure~\ref{fig5}(d1, d2, d3), the oxygen
abundance is used to represent the metallicity of star-forming
galaxies \citep{kauffmann03a}. The oxygen abundances of ISM were
provided by \citet{tremonti04}, which computed the abundances by
fitting serial spectral features from SDSS observational spectra to
models. The effect of AGN activities is shown in
Figure~\ref{fig5}(e1, e2, e3) by using $d_{AGN}$ to represent AGN
activities as like \citet{wu07}. Only AGNs located in between the
two boundaries defined by \citet{wu07} were adopted. We should
note, except the concentration parameters, all above factors are
derived from the spectral features in central regions covered by
SDSS fibers with 3$\arcsec$ diameter.
For the correlations between M/L(3.6$\mu$m) and the five factors (EW(H$\alpha$), $R$50/$R$90, E(B-V), oxygen abundance, $d_{AGN}$), the Spearman Rank-order
correlation analysis coefficients are 0.74, 0.58, 0.15, 0.45 and 0.01, respectively; while for the correlations between M/L(4.5$\mu$m)
and the five factors (bottom five panels in Figure~\ref{fig5}),the Spearman Rank-order correlation analysis coefficients are 0.77, 0.58, 0.10, 0.36 and 0.01, respectively.
Hence, it is clear that the M/L(3.6$\mu$m) and M/L(4.5$\mu$m) are more sensitive to the EW(H$\alpha$), morphology and metallicity of galaxies.
The value of the EW(H$\alpha$) represents SFH for galaxies without strong nuclear activities, while the morphology and metallicity are sensitive to the nowaday SFR other than the SFH.

In order to derive the formulae to estimate stellar masses using these three band luminosities, we present the correlations between
them and stellar masses for all the galaxies in our sample firstly. The $K$$_{\rm s}$ band, 3.6$\mu$m and 4.5$\mu$m bands luminosities
are plotted against stellar masses in Figure~\ref{fig6}. The stellar masses are the referenced one described in $\S$2.2. These galaxies
show good correlations between the three band luminosities and stellar masses. The best nonlinear and linear fits are shown as solid
and dotted lines in Figure~\ref{fig6}, and the fitting parameters are listed in Table 1. Here, the best nonlinear fits
are obtained by using two-variable regression. Similar to Li07, we can derive star mass formulae based on the non-linear correlations:
\begin{equation}
Log_{10} {~M_{\star,ref}}=(-1.60\pm0.05)+(1.12\pm0.02)\times Log_{10} { ~\nu L_{\nu}{[K_{\rm s}]}}
\end{equation}
\begin{equation}
Log_{10} {~M_{\star,ref}}=(-0.79\pm0.03)+(1.19\pm0.01)\times Log_{10} { ~\nu L_{\nu}{[3.6\mu m]}}
\end{equation}
\begin{equation}
Log_{10} {~M_{\star,ref}}=(-0.25\pm0.03)+(1.15\pm0.01)\times Log_{10} { ~\nu L_{\nu}{[4.5\mu m]}}
\end{equation}
%
The correlations between two IRAC band luminosities and referenced stellar mass are as tight as that between the $K$$_{\rm s}$ band luminosities and referenced stellar masses.
The fitting residuals' standard deviations and the Spearman Rank-order correlation analysis coefficients of the correlations of 3.6$\mu$m and 4.5$\mu$m luminosities with stellar masses are 0.11 and 0.12, 0.96 and 0.95, respectively.
The dashed line shown in Figure~\ref{fig6}(b) represents the non-linear fits for local luminous galaxies in the northern $Spitzer$
SWIRE fields obtained by \citet{li07}. Comparing the best non-linear fits in this work with that of \citet{li07}, an obvious downward
shift exists, especially for the non-luminous ones. For example, for galaxies with 3.6$\mu$m luminosity of $~10^{9}L_{\odot}$,
the stellar mass estimated from formula (2) would be about 0.5~dex lower than the result from corresponding formula in \citet{li07}.
The shift could be due to fact that different stellar mass references (Figure~\ref{fig2}) are used in these two works.


Above analysis ignored the effect by different SFH. In fact, there is such effect, at least for star-forming galaxies illustrated
in Figure~\ref{fig5}(a1, a2, a3), which showed the correlations between the EW(H$\alpha$) and mass-to-luminosity
ratios of $K$$_{\rm s}$ band, 3.6$\mu$m and 4.5$\mu$m luminosities. The SFH could also be represented with the $g$-$r$ color indicated by both \citet{bell01} and \citet{kauffmann03a}, which was also proved
an effective method to estimate the stellar mass by \citet{gallazzi09}. Figure~\ref{fig7} shows the correlation between $g$-$r$
color and the EW(H$\alpha$) of our sample galaxies. There exists obvious anti-correlation between -EW(H$\alpha$) and $g$-$r$ color.
Therefore, we calibrate the correlations between mass-to-luminosity ratios and $g$-$r$ color. Panels in Figure~\ref{fig8} show the correlations between various mass-to-luminosity ratios and $g$-$r$ colors.
Here, SDSS model magnitudes are used, and the foreground extinction and redshift have been corrected.

These mass-to-luminosity ratios that are shown in Figure~\ref{fig8} are not constantly following the changing of color. From these four
panels, we find out that redder galaxies tend to have larger mass-to-luminosity ratios. Using the two-variable regression, we obtain
the best non-linear fits illustrated as solid lines in four panels of Figure~\ref{fig8} and the fitting parameters are listed
in Table 2. According to these correlations, we derive the formulae to estimate the stellar mass:
\begin{equation}
Log_{10} \frac{M_{\star,ref}}{\nu L_{\nu}(r)} = (-0.81\pm0.01) + (1.47\pm0.01)\times (g-r)
\end{equation}
\begin{equation}
Log_{10} \frac{M_{\star,ref}}{\nu L_{\nu}(K_{\rm s})} = (-1.29\pm0.05) + (1.42\pm0.06)\times (g-r)
\end{equation}
\begin{equation}
Log_{10} \frac{M_{\star,ref}}{\nu L_{\nu}(3.6)} = (0.23\pm0.01) + (1.14\pm0.01)\times (g-r)
\end{equation}
\begin{equation}
Log_{10} \frac{M_{\star,ref}}{\nu L_{\nu}(4.5)} = (0.39\pm0.01) + (1.17\pm0.02)\times (g-r)
\end{equation}
The dotted-dashed line in Figure~\ref{fig8}(a) is the respective
line by \citet{bell03} (see their Table.7, Fig.6 and Fig.20). The
slope of our fit of M/L($r$) vs. $g$-$r$ color is much deeper than
that of \citet{bell03}. \citet{gallazzi09} also found similar
discrepancy. One possible explanation is that different
evolutionary population synthesis models have been used to derive
standard stellar mass. The correlation between M/L($r$) and
$g$-$r$ color is the tightest one, with the Spearman Rank-order
correlation analysis coefficients of 0.97 and the fitting
residuals' standard deviations of ~ 0.03. The correlations of
M/L(3.6$\mu$m) and M/L(4.5$\mu$m) with $g$-$r$ color are also
tight, with fitting residuals' standard deviations of ~ 0.05 and
0.06, and the Spearman Rank-order correlation analysis
coefficients of 0.83 and 0.80, respectively. The scatter in
Figure~\ref{fig8}(a) is apparently much smaller than those in
other panels, but the slope of the correlation between M/L($r$)
and $g$-$r$ color for red galaxies is a little shallower than the
slope for blue ones. This variation is also suggested in
\citet{kauffmann03a} based on about 100,000 SDSS galaxies. Besides
SFH, some other factors could possibly account for this slope
variation, too. This variation can not be found in other panels in
Figure~\ref{fig8}. If the variations exist in Figure~\ref{fig8}(b,
c, d), we can not clearly see them due to large scatters. The
distribution of mass ratios of derived stellar mass and the
reference one in Figure~\ref{fig9} show that the stellar masses
derived by using Eq.6 and 7 are much tighter than the respective
ones from Eq.2 and 3.

\section{Discussion}

\subsection{Comparion with $K$$_{\rm s}$ band}

$K$$_{\rm s}$ band luminosity was often used to derive stellar masses of galaxies. From Figure~\ref{fig5}, Figure~\ref{fig6} and
Figure~\ref{fig8}, we find that IRAC 3.6$\mu$m and 4.5$\mu$m luminosities are also good stellar mass tracers compared with $K$$_{\rm s}$
band luminosity, because the radiation in all these bands was dominated by old stellar populations. Additionally, for the same
sky area, the effect of extinctions in IRAC 3.6$\mu$m and 4.5$\mu$m photometries is absolutely weaker than that in $2MASS$ $K$$_{\rm s}$
band \citep[e.g.,][]{gao09}. There has been very few calibrations between stellar masses and 3$\mu$m to 5$\mu$m emissions. One
major reason is the limit of the weak transmissions of earth's atmosphere in this wavelength range. Now, the progress of space
astronomy make us capable to observe the MIR sky directly.

\citet{bell03} indicated that the dispersion of blue galaxies was
diffused in the M/L($K$$_{\rm s}$) vs. $B$-$R$ plane (their Figure
20). And the fitting slope was shallower for red galaxies compared
to the blue ones. Using modified broad band Johnson photometry
correlation, \citet{kannappan07} also found similar variation of
the slope. The transformation point of the slope of
\citet{kannappan07} was $1.2$ in Johnson optical color $B$-$R$,
which corresponded to a SDSS $g$-$r$ color of 0.55 based on the
transformation methods supplied by \citet{fukugita96}. In their
work, the slope of the correlation of M/L($K$$_{\rm s}$) vs.
$B$-$R$ was 0.5 for blue galaxies with $B$-$R$ color less than
1.2, while for the red galaxies with $B$-
the corresponding slope was 0.34. \citet{bell03} found out that
the reason for diffused dispersion and transformation of slope was
the difference of metallicity of blue galaxies. For two galaxies
with same in M/L($K$$_{\rm s}$), the one with lower metallicity
must be bluer than the higher metallicity one  optical colors.
However we do not find such phenomena in other three panels(
Figure~\ref{fig8}(b,c,d)), because of the lack of low metallicity
galaxies in this sample, or the larger dispersion.
Therefore, it must be more careful when estimating stellar masses of low
metallicity galaxies using NIR or MIR luminosities. We will investigate their properties in more details in future.

\subsection{Effect of Star Formation}

Besides the EW(H$\alpha$), mass-to-luminosity ratios of two MIR
luminosities are also sensitive to the concentration parameters
which could be seen in Figure~\ref{fig5}. Figure~\ref{fig10} shows
the correlations between stellar mass ratios (the stellar mass
derived from above formulae divided by the reference stellar mass)
and morphology ($R$50/$R$90). The derived stellar masses in
Figure~\ref{fig10}(a, b) base on nonlinear fits of 3.6$\mu$m and
4.5$\mu$m luminosities (with superscript 'A') vs. reference
stellar masses (Eq.2 and 3); while those in Figure~\ref{fig10}(c,
d) base on nonlinear fits of M/L(3.6$\mu$m) and M/L(4.5$\mu$m) vs.
$g$-$r$ color (Eq.6 and 7) (with superscript 'B'). We find out
that the obvious discrepancies of derived stellar masses of
galaxies with different morphologies in Figure~\ref{fig10}(a, b)
would overestimate the stellar masses of the late-type galaxies
compared to the early-type ones; while the ratios keep nearly
constant in Figure~\ref{fig10}(c, d). SFH could account for the
discrepancies in Figure~\ref{fig10}(a, b) when comparing the
slopes of two upper panels and two bottom panels. Except for SFH,
SFR may be another factor resulting in overestimations of stellar
masses for late-type galaxies. Galaxies with different
morphologies have different SFRs. There are more new stars born in
late-type galaxies than in early-type ones \citep{li07}. SFR could
relate with the values of intrinsic extinction and metallicity as
shown in Figure~\ref{fig5}. Hence, it is necessary to check SFR on
our stellar mass formulae.

Panels in Figure~\ref{fig11} are similar to those in
Figure~\ref{fig10}, just replacing $r$ band concentration
parameters by 24$\mu$m luminosities of galaxies. Same as
Figure~\ref{fig10}, stellar masses in two upper panels
(Figure~\ref{fig11}(a, b)) are derived by nonlinear fits of
3.6$\mu$m and 4.5$\mu$m luminosities (Eq.2 and 3), while those in
bottom panels (Figure~\ref{fig11}(c, d)) based on the nonlinear
fits of M/L(3.6$\mu$m) and M/L(4.5$\mu$m) vs. $g$-$r$ color (Eq.6
and 7). A template SED of a normal HII galaxy NGC~3351 (from
SINGS; \citep{kennicutt03}) was used to perform K-correction for
the 24$\mu$m band for all the sample galaxies with 24$\mu$m
detection. Figure~\ref{fig11} shows that Eq.2 and 3 would
overestimate the stellar mass for galaxies with higher 24$\mu$m
luminosities. The 24$\mu$m luminosity was proved to be a good SFR
tracer by $Spitzer$ observations \citep{wu05, calzetti05,
calzetti07, zhu08}. Therefore, Eq.2 and 3 are available for
galaxies with lower star-forming activities. While
Figure~\ref{fig11}(c, d) are almost flat. Thus, the Eq.6 and 7 are
not quite sensitive to SFR or morphology (Figure~\ref{fig10}).

In Figure~\ref{fig11}, there are three outliers which are marked with red crosses. The one with number '3' is an AGN, which will be discussed in next subsection.
Other two sources are star-forming galaxies with apparent features of interactions or mergers in SDSS images.
The one with number '1' is a dwarf star-forming galaxies, with low metallicity of 7.93 and very large 24$\mu$m-to-3.6$\mu$m flux ratio.
The one with number '2' is a luminous star-forming galaxies with strong PAH emissions in IRAC 8$\mu$m band.
Both of them have very blue $g$-$r$ color. Actually, they are the only two galaxies with $g$-$r$ color less than $0.1$ in Figure~\ref{fig8}.
Additionally, the stellar mass of '2' derived by using SDSS photometries and spectral indices are apparently different.
Therefore, it is hard to define which one is the best estimator.
The stellar mass estimated using photometries is about 1 dex less than the one estimated by spectral indices. For '1' and '3', there are no stellar masses values derived by spectral indices.

Apart from emission directly from the photosphere of old red
stars, the continuum from very small dust heated by hot stars,
even some broad emission features \citep{gillett73, willner77}
which were indicated to be from PAHs \citep{wen07, cao07}, could
increase the MIR luminosity of a galaxy. The dust emission
dominates radiation of star-forming galaxies in the wavelength
range beyond 5$\mu$m, but could also enhance the 4.5$\mu$m
flux\citep{helou04}. To evaluate such effect, the
4.5$\mu$m-to-3.6$\mu$m flux ratio is plotted against the 24$\mu$m
luminosity in Figure~\ref{fig12}. The three outliers in
Figure~\ref{fig11} were marked with red crosses. The blue
boxes are the AGNs. There are some AGNs with very high
4.5$\mu$m-to-3.6$\mu$m luminosity ratios marked with red boxes.
Except these outliers, we find a correlation between
4.5$\mu$m-to-3.6$\mu$m luminosity ratio and star formation for
most of sources. Because the 24$\mu$m luminosity is the tracer of
SFR, this correlation may be due to the dust emissions in the
4.5$\mu$m band. With the assumption that the entire IRAC 3.6$\mu$m
band luminosity is from the stellar emission, a factor of 0.596
\citep{helou04} was used to scale the stellar continuum of
3.6$\mu$m to that of 4.5$\mu$m based on {\sl Starburst99}
synthesis model \citep{leitherer99} with the solar metallicity and
a Salpeter initial mass function (IMF) between 0.1 and 120 $M_{\rm
\odot}$. This factor is similar to that derived from the observed
SED of early type galaxies by \citet{wu05}. The factor of 0.596 is
plotted with dashed line in Figure~\ref{fig12}. If this factor is
available for all star-forming galaxies, we could estimate the
dust emission fraction based on the correlations between 24$\mu$m
and total-infrared (TIR) luminosities\citep{zhu08}, e.g., for
luminous infrared galaxies with TIR luminosity of
$10^{11}L_{\odot}$, about ~30\% flux at rest-frame 4.5$\mu$m is
from dust emission. Therefore, compared to the 4.5$\mu$m emission,
IRAC 3.6$\mu$m emission is a better stellar mass tracer. This is
also supported by the much tighter correlation between 3.6$\mu$m
and stellar masses in Table 1 and 2.

\subsection{Effect of AGNs}

Due to the powerful UV emissions from AGNs, the dust surrounding them could be heated
by central monsters and then re-radiate in IR range \citep[see, e.g,][]{bell05,perez06,elbaz07,daddi07}.
Higher MIR-to-H$\alpha$/MIR-to-UV luminosity ratios were found by previous works based
on $Spitzer$ observations \citep{wu07,li07,zhu08}. So we want to know whether the energy
from AGNs could strengthen the 3.6$\mu$m and 4.5$\mu$m emissions. Figure~\ref{fig13}
shows the correlations between stellar mass ratios
and AGN activities. Here, just the AGNs located inside the two boundaries defined by \citet{wu07}
are adopted. \citet{kauffmann03b} found that the contribution from AGN light to the optical continuum was much weaker compared
to from starburst. Thus, ignoring the effect of AGN emission on the estimation of referenced stellar mass would be reasonable.
Red boxes represent the respective AGNs in Figure~\ref{fig12} with 4.5$\mu$m-to-3.6$\mu$m flux ratio large than 0.9.
According to \citet{wen07}, we know that AGNs with very high 4.5$\mu$m-to-3.6$\mu$m ratios may be QSOs.
In Figure~\ref{fig13}, except those scatters, the stellar mass
derived from MIR luminosities keep constant with increasing $d_{AGN}$,
while Figure~\ref{fig5}(e2, e3) show that AGN activities can not disturb mass-to-luminosity ratio for AGN-hosting galaxies, even at 4.5$\mu$m(Figure~\ref{fig14}).
Therefore, there seems no confirmed non-thermal radiations or dust
emissions heated by AGNs in IRAC 3.6$\mu$m and 4.5$\mu$m
detections. Because all the sources here are type-II AGNs, the
powerful intrinsic extinction of dust torus of narrow line AGNs
could be another explanation for non-detection of AGNs in
3.6$\mu$m and 4.5$\mu$m in our sample. The absence of AGN
emissions at 3.6$\mu$m and 4.5$\mu$m also indicate that the
stellar mass formulae based on 3.6$\mu$m and 4.5$\mu$m
luminosities are applicate to most AGN-hosting galaxies.

Figure~\ref{fig4} present the distributions of absolute B-band magnitudes, reference stellar masses and $r$ band concentration parameters.
The distributions in Figure~\ref{fig4} of composite galaxies and AGNs are strikingly different from those of star-forming galaxies, but similar to absorption line galaxies.
These results indicate that AGNs in our sample tend to be hosted in early-type galaxies \citep{kauffmann03b, zhu08}.
Based on these results and the discussions about the effect of
star formation in $\S$4.2, we find star formation rates in our
AGN-hosting galaxies are lower.

\section{Summary}

Based on the sample cross-identified from Spitzer SWIRE field and
SDSS spectrographic survey, we derived the formulae to calculate
the stellar mass using non-linear correlations wth IRAC 3.6$\mu$
and 4.5$\mu$m luminosities. The mass-to-luminosity ratio of these
two MIR luminosities are sensitive to the EW(H$\alpha$),
morphology and metallicity of galaxies, especially for the former
one, while the value of the EW(H$\alpha$) represents the SFH for
normal galaxies. Using the $g$-$r$ color to represent SFH, we
re-calibrate the stellar mass formulae for various galaxies. We
find these formulae are better than those which ignore the effect
of SFH. We could not conclude the applicability to low metallicity
galaxies due to a lack of such object in our sample, thus one
must be really careful when adopting our formulae to estimate
stellar masses of those galaxies. Additionally, we found that the
dust emission heated by hottest young stars could enhance the IRAC
4.5 $\mu$m luminosity, especially for luminous infrared galaxies.
So the formulae we derived are not applicate to these galaxies.
And these formulae are also not applicate to the stronger AGNs,
such as QSOs, although almost all the AGN-hosting galaxies in our
sample present similar properties like absorption line galaxies
(early-type), with low SFRs.

\acknowledgments

We thank the anonymous referee for constructive comments and suggestions. We acknowledge Drs. X.-Y. Xia, C.-N. Hao, Z.-G. Deng,
Y. Gao, Q.-S. Gu, X.-Z. Zheng, J,-Z. Wang for advice and helpful discussions, and J.-S. Huang, Z. Wang, J.-L. Wang, and F.-S. Liu for their
capable help and assistance throughout the process of $Spitzer$ data reductions.

This project is supported by NSFC grants 10833006 and 10773014, and by the 973 Program grant 2007CB815406.

This work is based on observations made with the $Spitzer$ Space Telescope, which is operated by Jet Propulsion Laboratory of
the California Institute of Technology under NASA Contract 1407. We gratefully acknowledge University of Massachusetts and NASA/IPAC
support for supply $2MASS$ data. Funding for the SDSS and SDSS-II has been provided by the Alfred P. Sloan Foundation, the Participating
Institutions, the National Science Foundation, the U.S. Department of Energy, the National Aeronautics and Space Administration,
the Japanese Monbukagakusho, the Max Planck Society, and the Higher Education Funding Council for England. The SDSS Web Site is
http://www.sdss.org/. The SDSS is managed by the Astrophysical Research Consortium for the Participating Institutions. The Participating
Institutions are the American Museum of Natural History, Astrophysical Institute Potsdam, University of Basel, University of
Cambridge, Case Western Reserve University, University of Chicago, Drexel University, Fermilab, the Institute for
Advanced Study, the Japan Participation Group, Johns Hopkins University, the Joint Institute for Nuclear Astrophysics, the Kavli
Institute for Particle Astrophysics and Cosmology, the Korean Scientist Group, the Chinese Academy of Sciences (LAMOST), Los
Alamos National Laboratory, the Max-Planck-Institute for Astronomy (MPIA), the Max-Planck-Institute for Astrophysics (MPA), New
Mexico State University, Ohio State University, University of Pittsburgh, University of Portsmouth, Princeton University, the United States Naval Observatory, and the University of Washington.

\clearpage

\clearpage

\begin{figure}
\centering
\figurenum{2} \epsscale{} 
\caption{Comparing the referenced stellar mass, which derived by photometries fitting, with those from spectral features fitting
supplied by $MPA$/$JHU$ (Panel (a)) for about eighty thousand galaxies randomly selected from SDSS DR4. In Panel (b), the comparing
between the reference and the stellar mass computed based on the rest-frame $g$-$r$ color, $r$ band model magnitude and the equation
by \citet{bell03} is presented. Please note that a factor of 0.15 was adding to the equation of \citet{bell03} to correct the
'diet Salpeter' IMF to the normal Salpeter \citep{salpeter55} IMF.}
\label{fig2}
\end{figure}
\clearpage

\begin{figure}
\centering
\figurenum{3} \epsscale{} 
\caption{The correlations between the stellar mass ratio and the referenced stellar mass (a), $g$-$r$ color for about eighty thousand
galaxies randomly selected from SDSS DR4. The stellar mass ratio represent the discrepancies between the referenced stellar mass
and the stellar mass estimated based on the rest-frame $g$-$r$ color, $r$ band model magnitude and the equation by \citet{bell03}
is presented. Same as in Figure~\ref{fig2}, a factor of 0.15 was adding to the equation of \citet{bell03} to correct the 'diet
Salpeter' IMF to the normal Salpeter \citep{salpeter55} IMF.}
\label{fig3}
\end{figure}
\clearpage

\begin{figure}
\centering
\figurenum{4} \epsscale{} \plotone{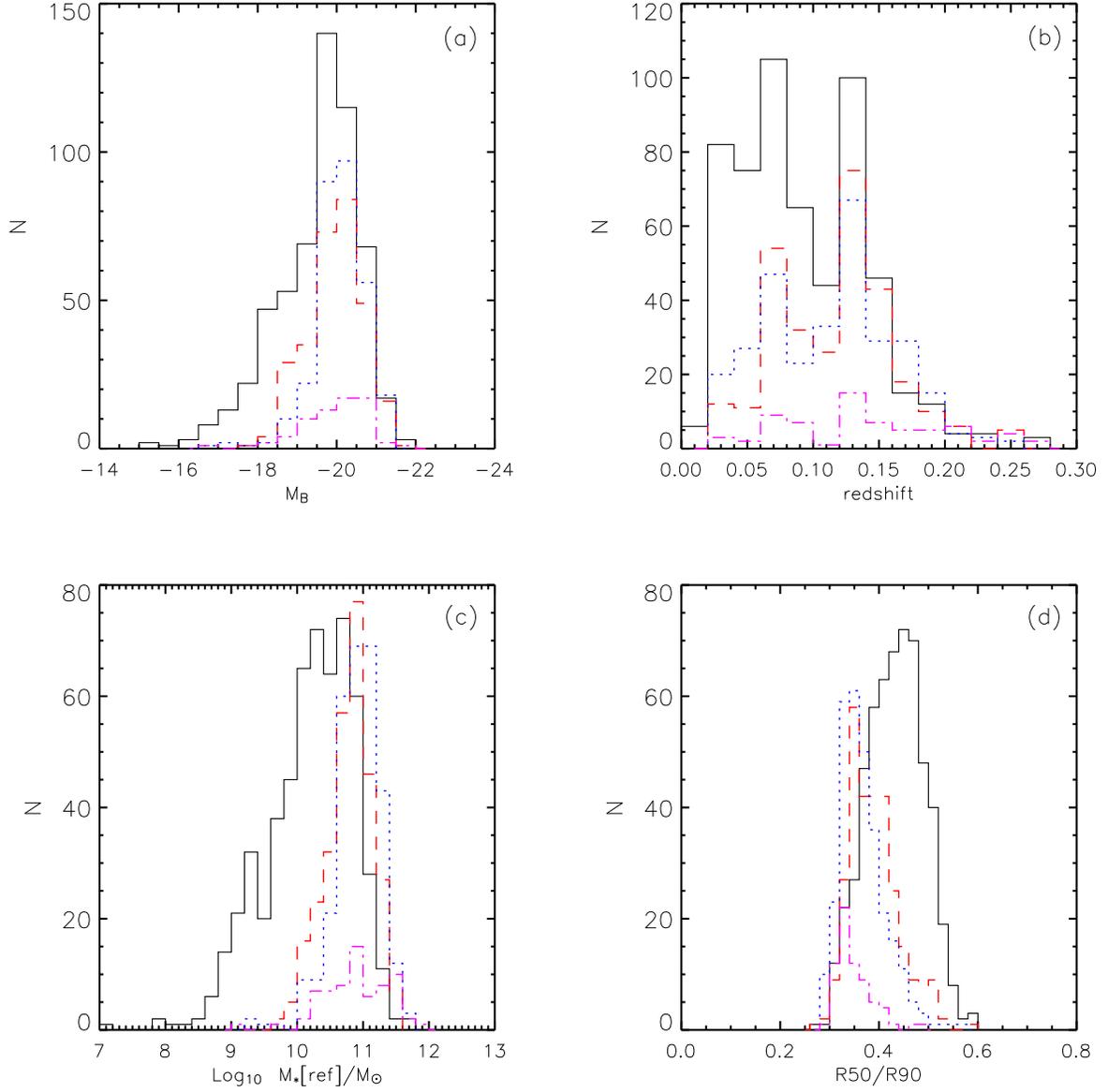}
\caption{The distributions of (a) absolute B-band magnitude; (b) redshift; (c) the referenced stellar mass; (d) the concentration
parameters ($R$50/$R$90) for star forming galaxies (black solid lines), composites (red dashed lines), AGNs (blue dotted lines)
and absorption line galaxies (pink dashed-dotted lines).}
\label{fig4}
\end{figure}
\clearpage

\begin{figure}
\centering
\figurenum{5} \epsscale{} \plotone{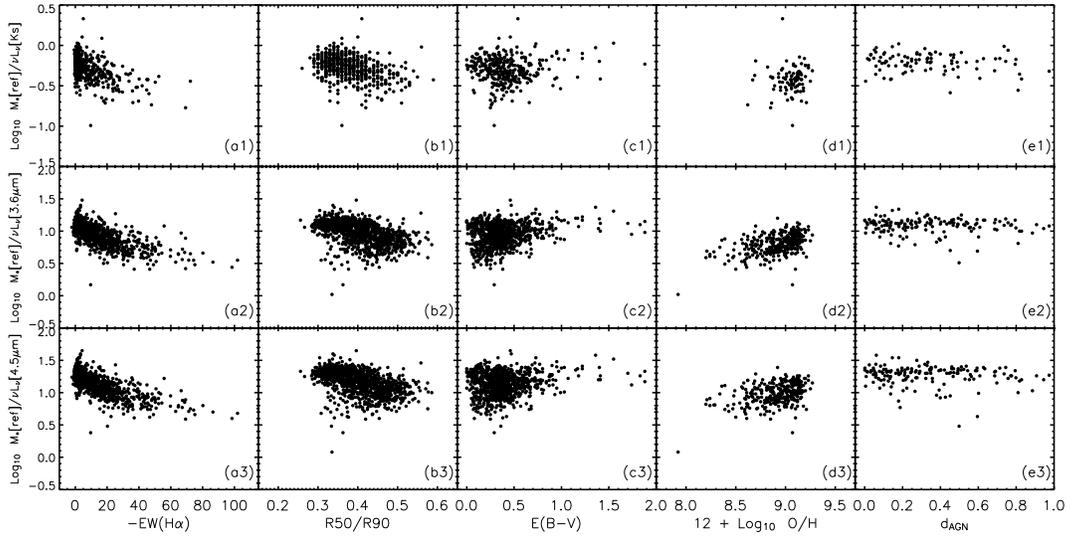}
\caption{Comparing the mass-to-luminosity ratios with various factors, include H$\alpha$ equivalent width (a1, a2, a3), $R$50/$R$90 (b1, b2, b3),
intrinsic reddening (c1, c2, c3), oxygen metallicity (d1, d2, d3) and the activity of AGN (e1, e2, e3).
The $2MASS$ $K$$_{\rm s}$ band luminosity is used in upper five panels. The 3.6$\mu$m luminosity is used in middle five panels.
The 4.5$\mu$m luminosity is used in the bottom five panels.}
\label{fig5}
\end{figure}

\clearpage

\begin{figure}
\figurenum{6} \epsscale{} \plotone{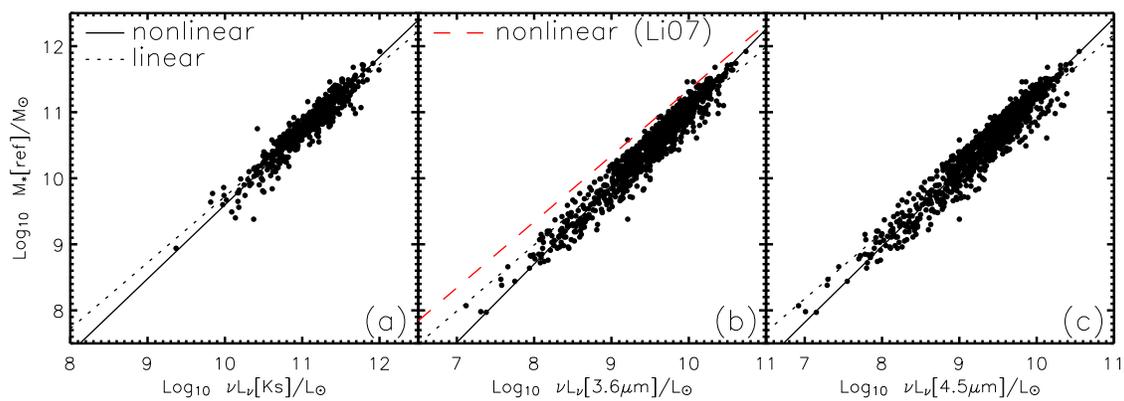}
\caption{Correlations between referenced stellar mass and $2MASS$ $K$$_{\rm s}$ band, two IRAC bands luminosities for all the galaxies.
The best nonlinear and linear fits are illustrated as solid and dotted lines. The red dashed lines show in Panels (b) represent
the nonlinear fits for more than one hundred luminous SWIRE-field galaxies by \citet{li07}.}
\label{fig6}
\end{figure}

\clearpage

\begin{figure}
\figurenum{7} \epsscale{} \plotone{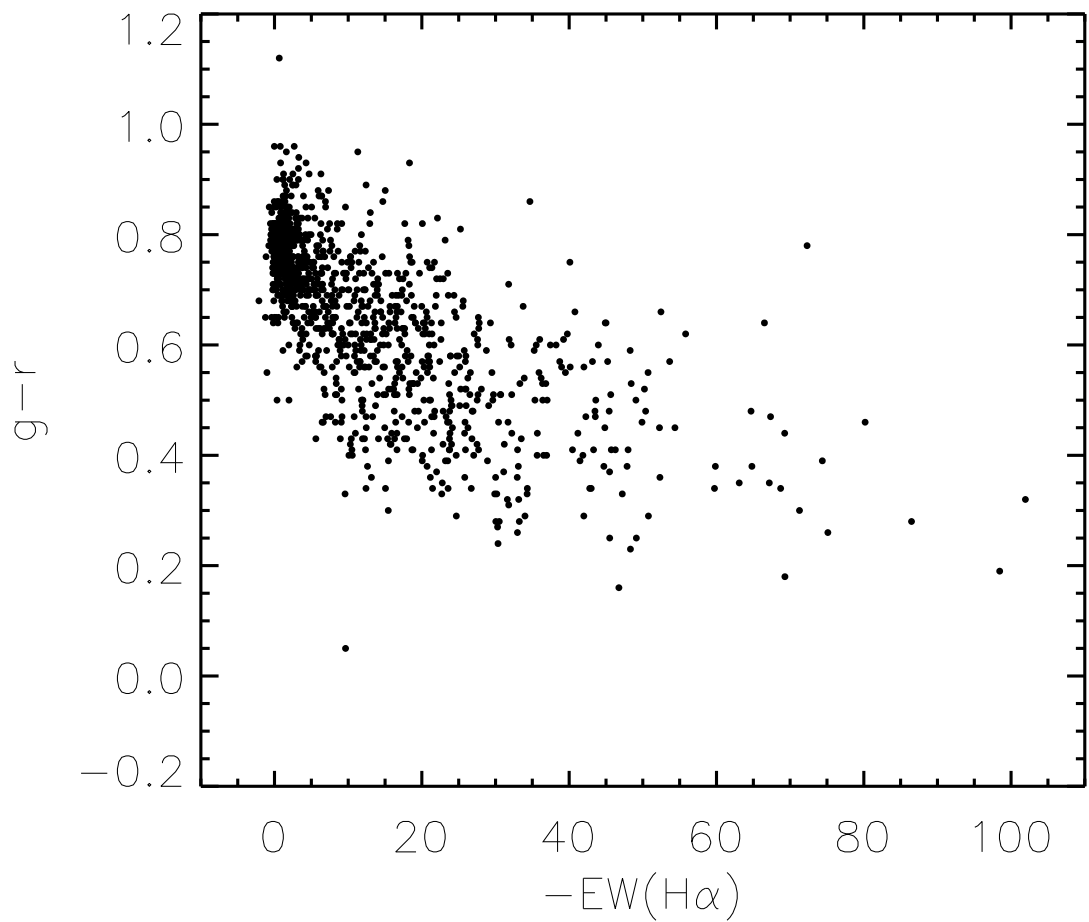}
\caption{Correlations between EW(H$\alpha$) and $g$-$r$ color for all the galaxies.}
\label{fig7}
\end{figure}

\begin{figure}
\figurenum{8} \epsscale{} \plotone{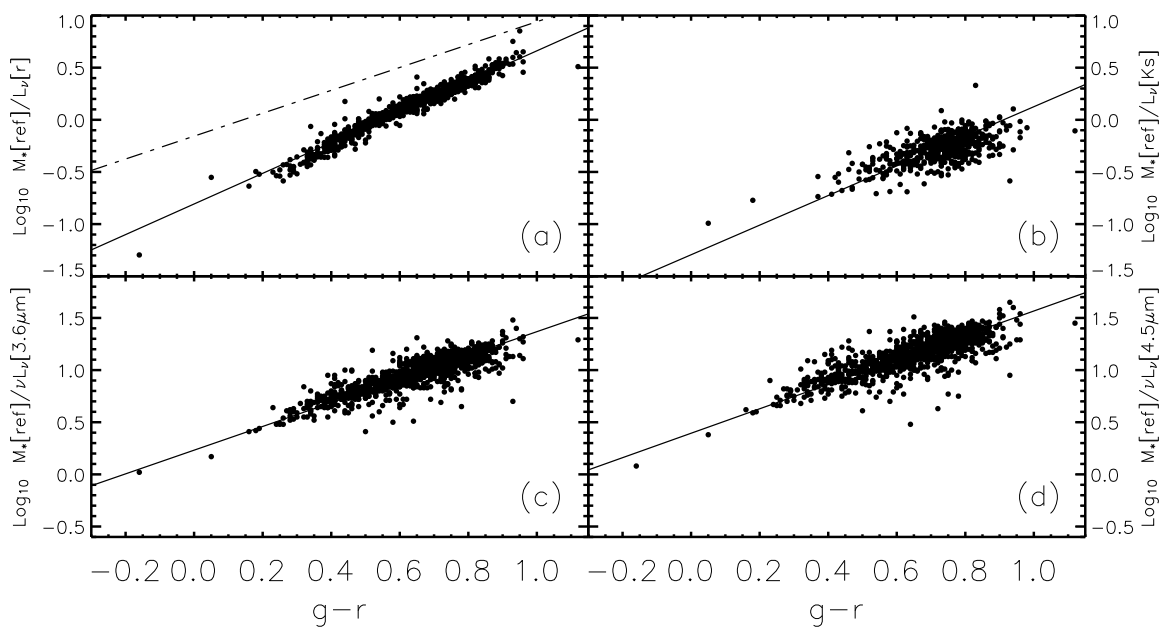}
\caption{Correlations between $g$-$r$ color and various mass-to-luminosity ratios for all the galaxies.  Due to the relatively
lower sensitivity in $2MASS$ $K$$_{\rm s}$ band observation, only 544 galaxies were detected and presented in Panel (b). The solid lines
represent the best nonlinear fits. The dotted-dashed line in Panel (a) is the respective line by \citet{bell03}.}
\label{fig8}
\end{figure}
\clearpage

\begin{figure}
\figurenum{9} \epsscale{} \plotone{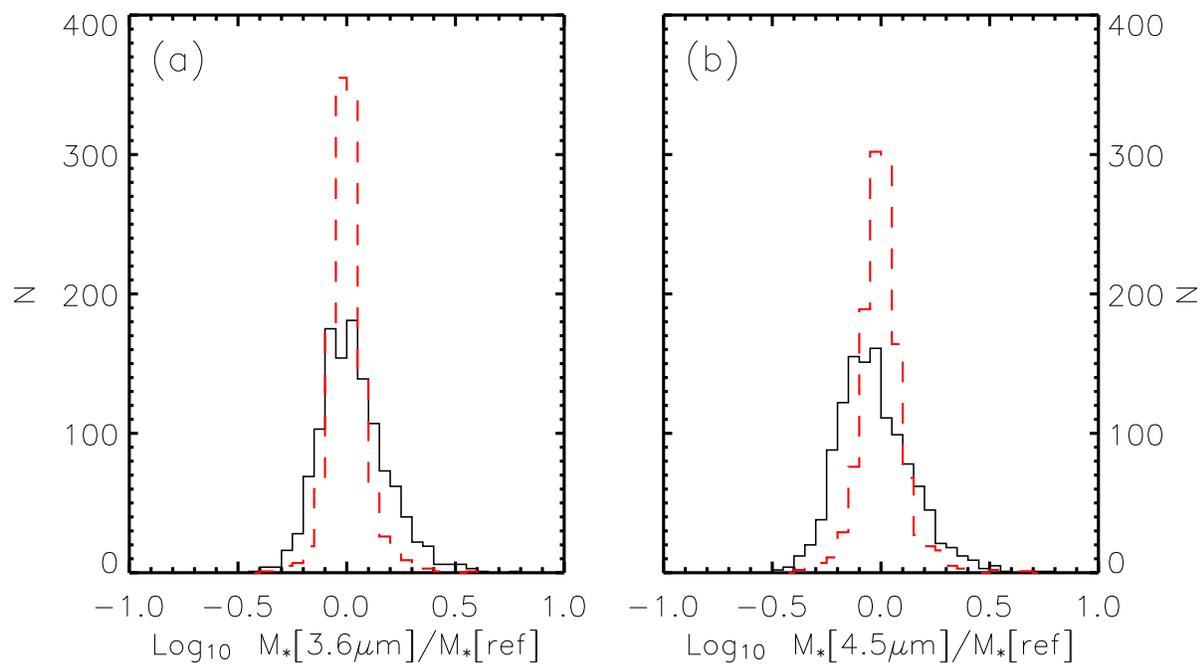}
\caption{The distributions of ratios of the derived stellar masses divided by the
referenced ones. The derived stellar masses were computed based on Eq.2 (black solid
lines in Panel (a)), Eq.6 (red dashed lines in Panel (a)), Eq.3 (black solid lines
in Panel (b)), Eq.7 (red dashed lines in Panel (b))}
\label{fig9}
\end{figure}
\clearpage

\begin{figure}
\figurenum{10} \epsscale{} \plotone{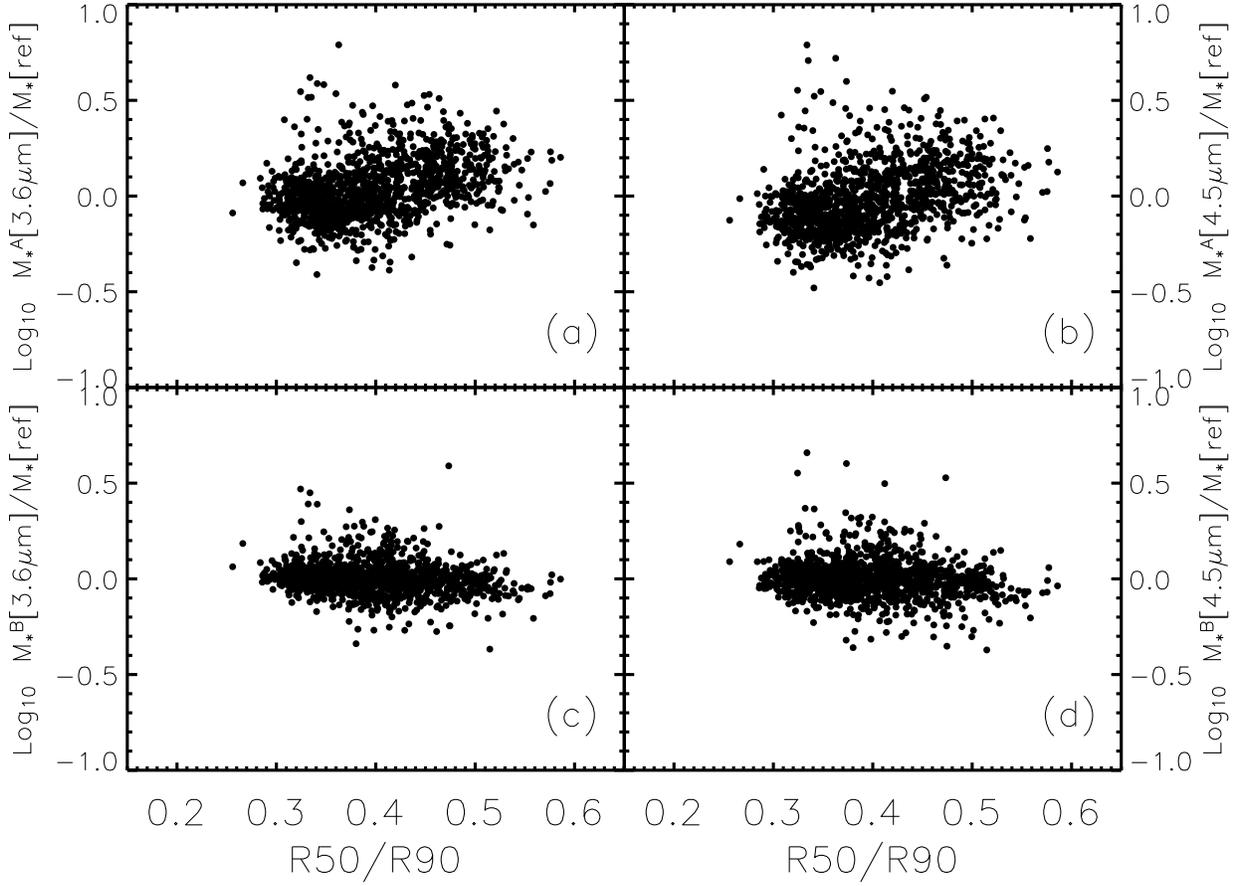}
\caption{Correlations between concentration parameters and the stellar mass ratios,
which are the discrepancies between the stellar mass derived from our calibrated
formulae and the reference. The derived stellar mass by using Eq.2, 3, 6, 7 is shown in Panel (a), (b), (c), (d), respectively.}
\label{fig10}
\end{figure}
\clearpage

\begin{figure}
\figurenum{11} \epsscale{} \plotone{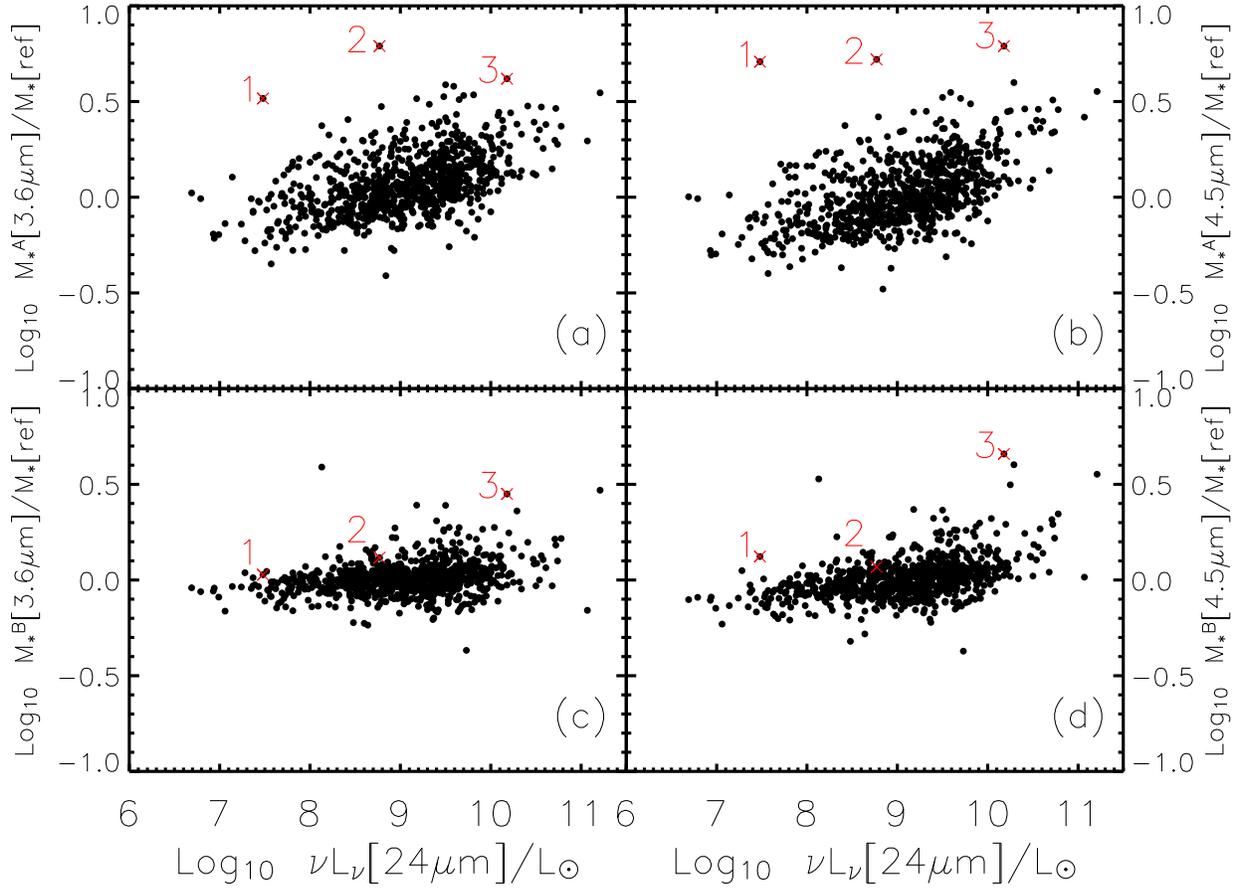}
\caption{
Similar as Figure 10, but the concentration parameters were instead of 24$\mu$m luminosities. The red crosses represent the three scatters, and they were signed out with numbers.
}
\label{fig11}
\end{figure}
\clearpage

\begin{figure}
\figurenum{12} \epsscale{} \plotone{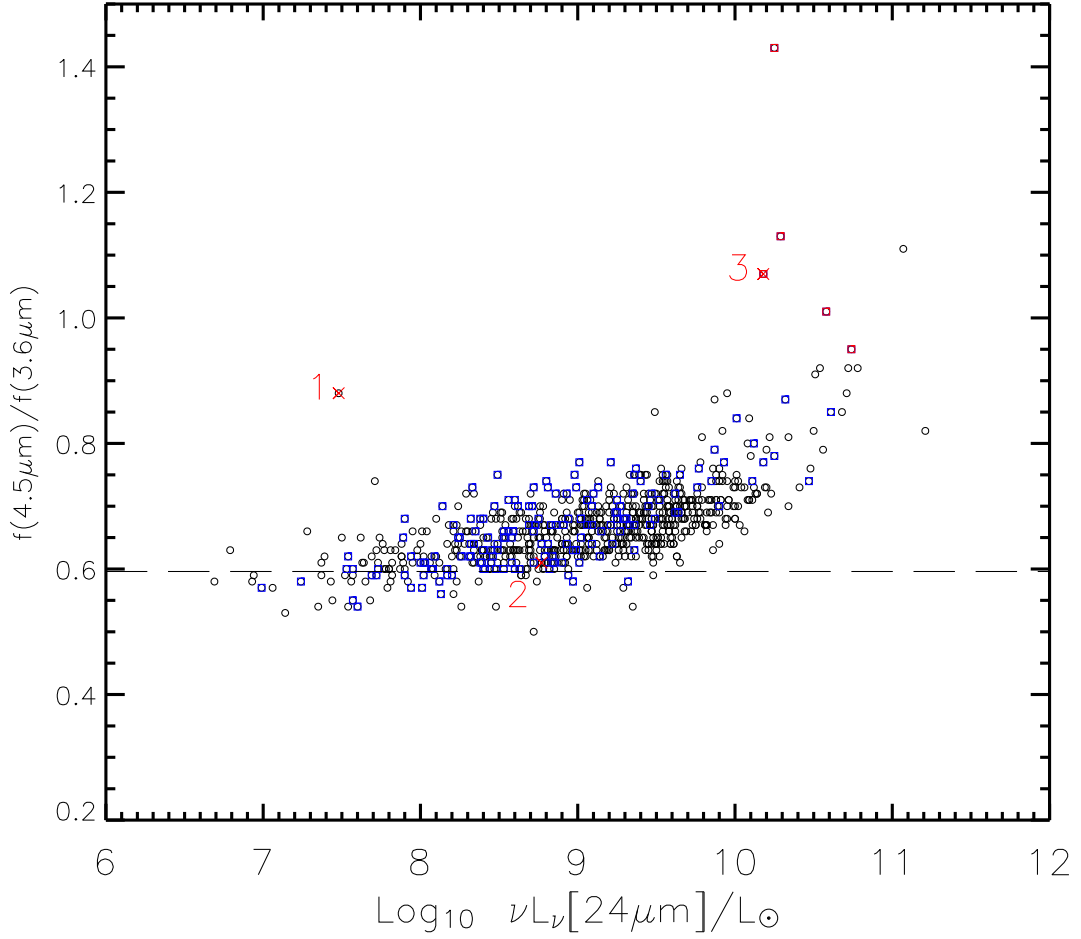}
\caption{Correlations between 24$\mu$m luminosity and the flux ratio between 3.6$\mu$m and 4.5$\mu$m for all the galaxies. The
blue boxes represent AGNs. The red boxes represent AGNs with 3.6$\mu$m-to-4.5$\mu$m ratio larger than 0.9. The three outliers in Figure 11 were sighed out with red crosses.
The dashed line represent the factor of 0.596 which is the ratio between 4.5$\mu$m and 3.6$\mu$m flux by \citep{helou04}
with the assumption that the entire IRAC 3.6$\mu$m and 4.5$\mu$m band luminosities are from the stellar emission.}
\label{fig12}
\end{figure}
\clearpage

\begin{figure}
\figurenum{13} \epsscale{} \plotone{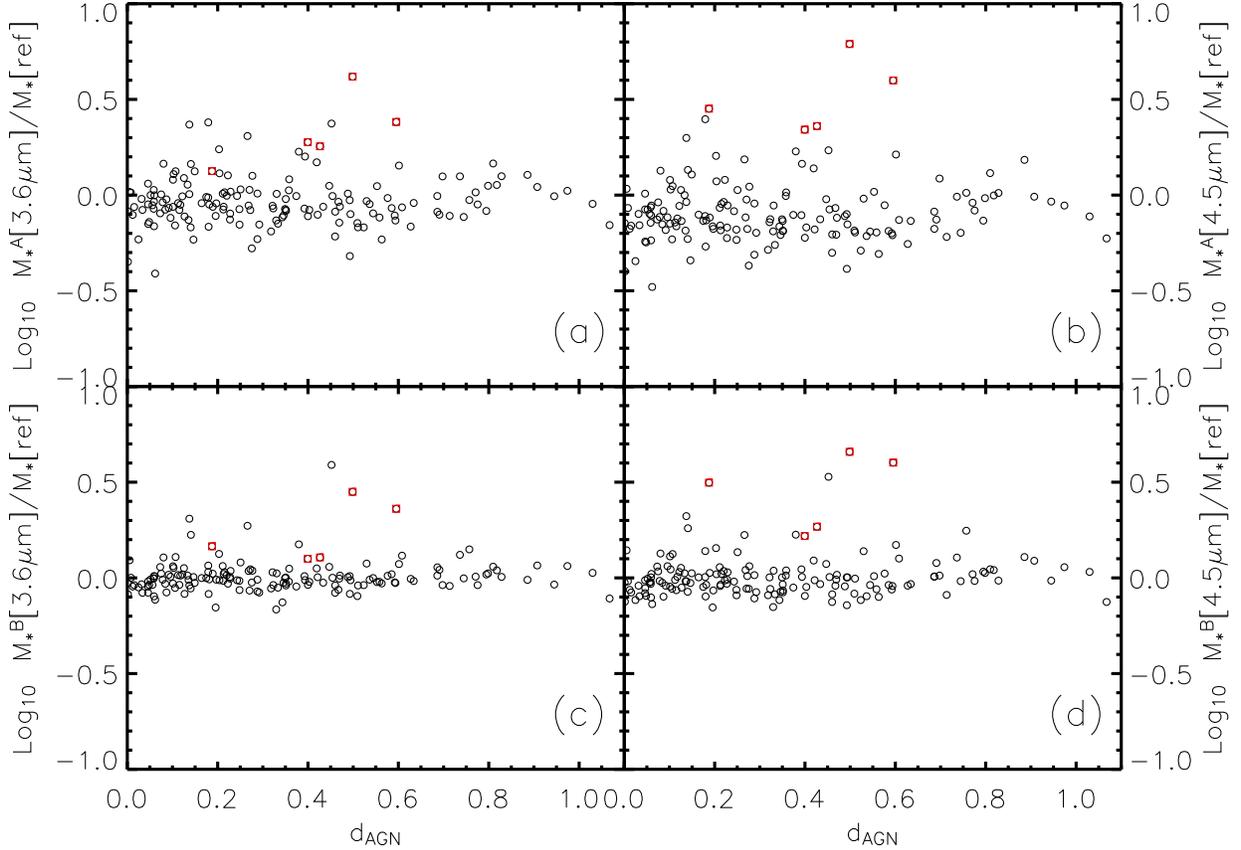}
\caption{Correlations between AGN activity and the stellar mass ratios, which are the discrepancies between the
stellar mass derived from our calibrated formulae and the reference. Here, the distance $d_{AGN}$, which was defined by \citet{wu07},
to characterize AGN activity. The derived stellar mass by using Eq.2, 3, 6, 7 is shown in Panel (a), (b), (c), (d), respectively.
The red boxes represent the respective AGNs in Figure~\ref{fig12} with higher 4.5$\mu$m-to-3.6$\mu$m flux ratio (large than 0.9).}
\label{fig13}
\end{figure}
\clearpage

\begin{figure}
\figurenum{14} \epsscale{} \plotone{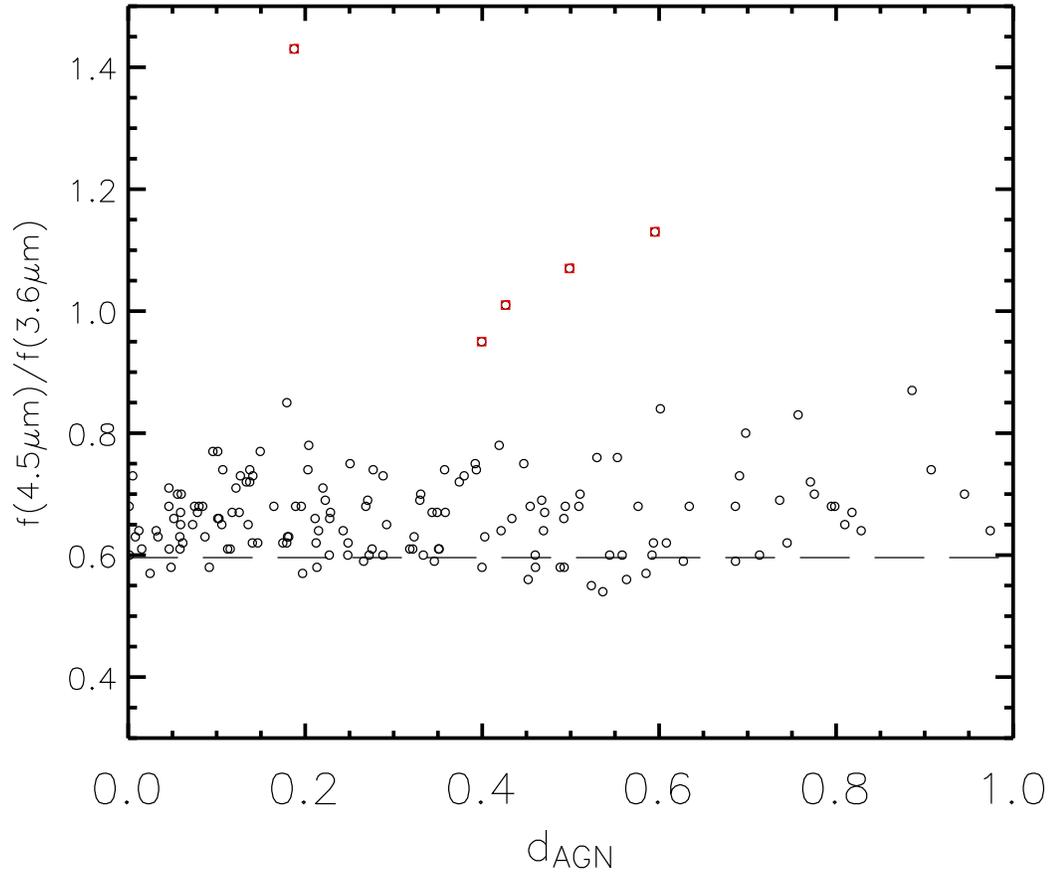}
\caption{Correlations between the distance $d_{AGN}$, which represent AGN activity, and the flux ratio between 3.6$\mu$m and 4.5$\mu$m for AGNs.
The red boxes represent the sources with 4.5$\mu$m-to-3.6$\mu$m flux ratio large than 0.9. The dashed line represent the factor of 0.596 such as plotted in Figure~\ref{fig12}.}
\label{fig14}
\end{figure}
\clearpage

%
\begin{deluxetable}{llrcccrr}
\centering
\tablecolumns{8}
\tabletypesize{\footnotesize}
\tablewidth{0pt}
\tablecaption{Correlations coefficients between stellar mass and $K$$_{\rm s}$ band, 3.6$\mu$m and 4.5$\mu$m luminosities}
\tablehead{
\colhead{$y$} & \colhead{$x$} & \colhead{$a$} & \colhead{$b$} &
\colhead{$s$} & \colhead{$r$} & \colhead{$c$} & \colhead{$N$}\\
\colhead{(1)} & \colhead{(2)} & \colhead{(3)} & \colhead{(4)} &
\colhead{(5)} & \colhead{(6)} & \colhead{(7)} & \colhead{(8)}}
\startdata
$ M_{\rm *} $ & $ \nu L_{\nu}[K_{\rm s}]$&$-$ 1.60$\pm$0.05& 1.12$\pm$0.02& 0.10& 0.94& 0.28$\pm$0.11& 544\\
$ M_{\rm *} $ & $ \nu L_{\nu}[3.6\mu m] $&$-$ 0.79$\pm$0.03& 1.19$\pm$0.01& 0.11& 0.96& 0.99$\pm$0.14& 1220\\
$ M_{\rm *} $ & $ \nu L_{\nu}[4.5\mu m] $&$-$ 0.25$\pm$0.03& 1.15$\pm$0.01& 0.12& 0.95& 1.17$\pm$0.14& 1220\\
\hline
\enddata
\tablecomments{Col.(1): the referenced stellar mass in solar unit; Col.(2): names of MIR luminosities;
Col.(3)-(4): the coefficients $a$ and $b$ of the nonlinear fit: $\log_{10}(y)=$a$+$b$\log_{10}(x)$;
Col.(5): the standard deviation $s$ of the fitting residuals;
Col.(6): the coefficient $r$ of the Spearman Rank-order correlation analysis;
Col.(7): the coefficient $c$ of the linear fit: $\log_{10}(y)=$c$+\log_{10}(x)$;
Col.(8): the number of sample galaxies used for the fitting procedures.}
\label{tab1}
\end{deluxetable}

\clearpage

%
\begin{deluxetable}{llrcccr}
\centering
\tablecolumns{7}
\tabletypesize{\footnotesize}
\tablewidth{0pt}
\tablecaption{Correlations of $M$$_{\rm*}$/$L$ and colors}
\tablehead{
\colhead{$y$} & \colhead{$x$} & \colhead{$a$} & \colhead{$b$} &
\colhead{$s$} & \colhead{$r$} & \colhead{$N$}\\
\colhead{(1)} & \colhead{(2)} & \colhead{(3)} & \colhead{(4)} &
\colhead{(5)} & \colhead{(6)} & \colhead{(7)} }
\startdata
$ M_{\rm *}/L_{\nu}(r)            $ & $ g-r $&$-$0.81$\pm$0.01& 1.47$\pm$0.01& 0.03& 0.97& 1215\\
$ M_{\rm *}/L_{\nu}(K_{\rm s})    $ & $ g-r $&$-$1.29$\pm$0.05& 1.42$\pm$0.06& 0.08& 0.54& 410\\
$ M_{\rm *}/\nu L_{\nu}[3.6\mu m] $ & $ g-r $&   0.23$\pm$0.01& 1.14$\pm$0.01& 0.05& 0.83& 1210\\
$ M_{\rm *}/\nu L_{\nu}[4.5\mu m] $ & $ g-r $&   0.39$\pm$0.01& 1.17$\pm$0.02& 0.06& 0.80& 1208\\
\hline
\enddata
\tablecomments{
Col.(1): various mass-to-luminosity ratios in solar unit; Col.(2): $g$-$r$ color;
Col.(3)-(4): the coefficients $a$ and $b$ of the nonlinear fit: $\log_{10}(y)=$a$+$b$\log_{10}(x)$;
Col.(5): the standard deviation $s$ of the fitting residuals;
Col.(6): the coefficient $r$ of the Spearman Rank-order correlation analysis;
Col.(7): the number of galaxies used for the fitting procedures.}
\label{tab2}
\end{deluxetable}

\clearpage

\end{document}